\documentclass[]{article}

\usepackage[text={15.5cm,23cm},centering]{geometry}
\usepackage[sectionbib]{chapterbib} 
\usepackage{aas_macros}
\usepackage{mathptmx}
\usepackage{graphicx}
\usepackage{subfigure}
\usepackage{color}
\usepackage{amssymb}
\usepackage{array}
\usepackage{threeparttable}
\usepackage{microtype}
\usepackage{natbib}

\newcommand{\chapterauthor}[1]{\textsc{#1}\section*{}}  

\DeclareUnicodeCharacter{2194}{\ensuremath{\leftrightarrow}}

\begin{document}

{\LARGE Formation of Planetesimals in the Outer Solar System} \\

\chapterauthor{Anders Johansen (University of Copenhagen), Michele T. Bannister (University of Canterbury), Luke Dones (SwRI), Seth Jacobson (Michigan State University), Kelsi Singer (SwRI), Kathryn Volk (Planetary Science Institute), Maria Womack (The U.S. National Science Foundation, University of Central Florida)}

\section{Abstract}

The Solar System hosts the most studied and best understood major and minor planetary bodies -- and the only extraterrestrial bodies to have been visited by spacecraft. The Solar System therefore provides important constraints on both the initial stages of planetary growth, communicated to us by its surviving planetesimal populations, and for the final result of the planet formation process represented by the architecture of the system and properties of the individual planets. We review here models of planetesimal formation in the outer Solar System as well as the wealth of recent observational constraints that has been used to formulate and refine modern planetesimal formation theory.


\bigskip
\section{A brief history of planetesimal formation models}

Observations and models of protoplanetary discs have given us a comprehensive insight into the growth of dust to pebbles and the radial drift of the pebbles. 
The planetesimal formation stage, in contrast, is hard to observe around other stars. 
The characteristic planetesimal size and birth-size distributions can in principle be inferred from dust production in debris discs, but the conclusions are highly dependent on the collisional cascade model and still give disparate answers \citep{Krivov+etal2018,KrivovWyatt2021}. 

The formation of planetesimals is a key step in the planet formation process. Historically, planetesimal formation theory was dominated by two major views, namely that planetesimals in the Solar System formed either by a gravitational instability of the sedimented mid-plane layer of grown dust aggregates \citep{Safronov1969,GoldreichWard1973} or by gradual coagulation from dust to kilometer-sized bodies \citep{Weidenschilling1980}. The gravitational instability model nevertheless requires unrealistically high densities to be reached by sedimentation to the mid-plane, while the coagulation model is hampered by the poor sticking of macroscopic dust aggregates \citep{DullemondDominik2005}. We therefore focus this chapter on reviewing newer work on how the streaming instability drives planetesimal formation by giving rise to large fluctuations in the mid-plane dust density that facilitate the gravitational contraction to form planetesimals.


\bigskip
\section {Overview of the Streaming Instability Model} 

The grand challenge for planetesimal formation models is to explain the growth from (sub-)micron-sized dust and ice particles to super-km-sized solid bodies (which we refer to as planetesimals). Reaching planetesimal sizes is an important accomplishment in planet formation, since the aerodynamical drag that causes mm-cm-sized pebbles to drift towards the star is strongly diminished when reaching size scales above a few hundred meters. The gas moves at slower than the Keplerian speed around the star, due to the outwards-declining pressure of the gaseous protoplanetary disc, causing dust particles to drift towards the star \citep{Weidenschilling1977}. The drift time-scale is proportional to the Stokes number of the particles, defined as ${\rm St} = \Omega t_{\rm s}$ where $\Omega$ is the Keplerian frequency at a given distance from the star and $t_{\rm s}$ is the aerodynamical stopping time of the particle to gas drag \citep{Johansen+etal2014}, when ${\rm St}\ll 1$. The Stokes number also determines the collision speed between particles induced by the turbulence \citep{OrmelCuzzi2007} and hence planetesimal formation theories are often formulated in terms of St rather than the particle size.

The protoplanetary disc that orbited our young Sun was endowed with an approximately 1\% mass fraction of dust and ice particles inherited from the interstellar medium. 
Experiments and modelling of collisions between dust aggregates have identified fragmentation as an important barrier to dust coagulation within the protoplanetary disc \citep{Chokshi+etal1993,Brauer+etal2008}. 
The collision speed, $v_{\rm c}$, between two equal-sized dust aggregates is given approximately by \citep{OrmelCuzzi2007,Birnstiel+etal2012}
\begin{equation}
  v_{\rm c} = \sqrt{3} \sqrt{\delta} \sqrt{\rm St} \, c_{\rm s}  \, ,
\end{equation}
where $\delta$ is a dimensionless number that quantifies the strength of the turbulence (which we use here as a diffusion analogy to the standard parameter $\alpha$ that describes viscosoty in accretion discs), the Stokes number ${\rm St}$ is (as defined above) a dimensionless number that characterizes the response time of the particles to gas motion and $c_{\rm s}=\sqrt{k_{\rm B} T/\mu}$ is the sound speed of the gas (here $k_{\rm B}$ is the Boltzmann constant, $T$ is the temperature and $\mu$ is the molecular mass of the gas particles). 
For both silicate particles and ice particles, the fragmentation threshold speed, $v_{\rm f}$, is likely approximately 1 m/s \citep{Guettler+etal2010,MusiolikWurm2019} This yields a maximum Stokes number of
\begin{equation}
  {\rm St} = \frac{1}{3} \delta^{-1} \left( \frac{v_{\rm f}}{c_{\rm s}} \right)^2 = 0.01 \left( \frac{\delta}{10^{-4}} \right)^{-1} \left( \frac{T}{100\,{\rm K}} \right)^{-1} \left( \frac{v_{\rm f}}{1\,{\rm m\,s^{-1}}} \right)^2
  \label{eq:St}
\end{equation}
This value of St corresponded to dust aggregates between 0.1 and 1 mm in size in the trans-Neptunian region of the solar protoplanetary disc \citep{Johansen+etal2014}. 
The fragmentation barrier may be circumvented if the dust monomers, the individual building blocks of dust aggregates, are small (sub-micron) and dominated by sticky ice \citep{Okuzumi+etal2012}, but this contrasts with the dominantly micron-sized matrix particles found in meteorites (\citealt{vanKooten+etal2019}; if the components found within meteorites are indeed representative of the outer Solar System) as well as experiments showing that ice particles at low temperatures stick no better than silicates \citep{MusiolikWurm2019}. More volatile ices irradiated by UV in the outer disc may nevertheless have higher sticking thresholds \citep{Musiolik2021}, so we emphasize that the exact value of the critical fragmentation speed of dust/ice aggregates in protoplanetary discs is far from settled.

The limited Stokes number reached at the fragmentation limit (equation \ref{eq:St}) implies that planetesimal formation must most likely proceed beyond pebble sizes with some aid of gravity. 
For gravity to become important, dense clumps of pebbles must first emerge within the gas. Many mechanisms have been proposed for such pebble concentration \citep{Johansen+etal2014}. 
Particles can concentrate passively within gaseous vortices and in pressure bumps \citep{BargeSommeria1995,KlahrBodenheimer2003,Johansen+etal2009b,Riols+etal2020}. It has also been shown experimentally and numerically that turbulent eddies at the smallest scales of a turbulent flow can concentrate dust particles between the eddies \citep{Cuzzi+etal2008}, but this mechanism does not appear to be efficient at concentrating pebbles in large enough amounts to drive gravitational collapse \citep{Pan+etal2011}. 
This small-scale particle concentration mechanism may also effectively become similar to trapping in vortices and pressure bumps when applied to larger scales of the turbulent flow \citep{Hopkins2016,Hartlep+etal2020}.

\begin{figure}
  \begin{center}
    \includegraphics[width=1.0\linewidth]{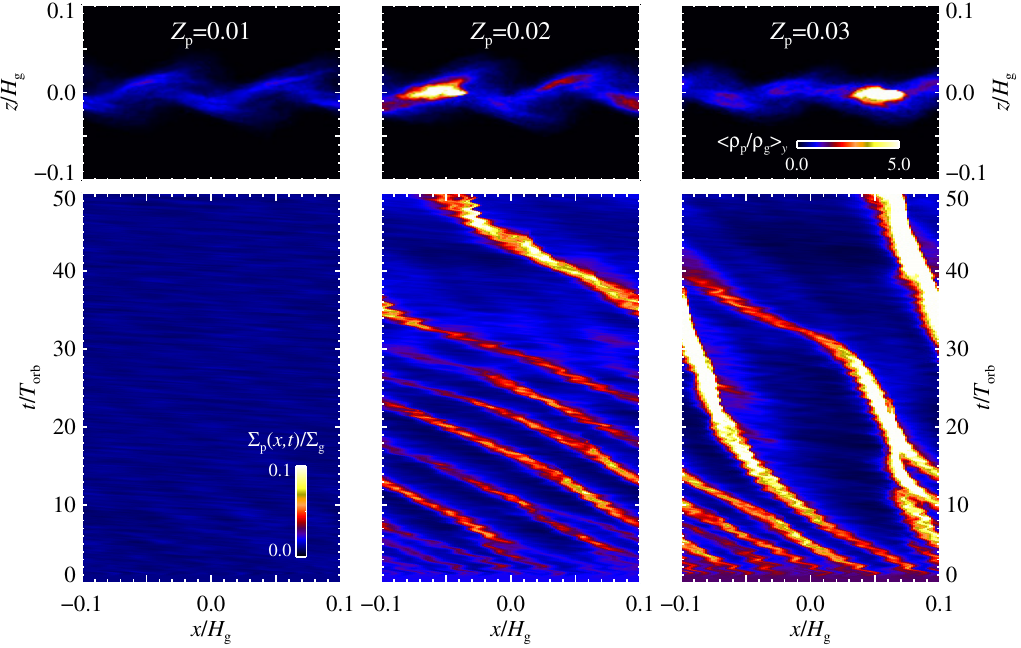}
  \end{center}
  \caption{Illustration of the formation of dense pebble filaments by the streaming instability, from \citep{Johansen+etal2009b}. The top panels show the density of pebbles in the mid-plane at the end of computer simulations that span a time-scale of 50 orbits ($x$ represents the radial direction in the disc and $z$ the vertical direction; $H_g$ is the scale height of the gas), while the bottom plots show space-time plots of the pebble surface density. The three columns display results for three values of the pebbles-to-gas surface density ratio $Z_{\rm p}$. The emergence of pebble-trapping filaments is clear for the two high-metallicity cases. These filaments move more slowly towards the star and thereby grow in mass by capturing free-drifting pebbles.}
  \label{f:Sigmapmx_t}
\end{figure}

The streaming instability is a physical mechanism that is driven by the radial drift of the particles toward the star. 
This drift is, as discussed above, driven by the headwind from the gas, which moves slightly slower than the Keplerian speed due to its small radial pressure support. 
The particles concentrate actively in the gas flow by exerting a frictional force onto the gas, which accelerates the gas towards the Keplerian speed and thereby reduces the headwind on the particles \citep{YoudinGoodman2005,JohansenYoudin2007,BaiStone2010}. This causes pebbles to spontaneously pile up into dense filaments where the headwind speed of the gas is low. This emergence of dense filaments in computer simulations of the streaming instability is illustrated in Figure \ref{f:Sigmapmx_t}.
The streaming instability in principle has positive growth rates for any combination of local dust-to-gas ratio $\epsilon$ and particle size parameterized through the Stokes number ${\rm St}$. 
The regime of $\epsilon \ll 1$ is nevertheless strongly affected by the size distribution of the particles, with analytical growth rates reduced by orders of magnitude when considering a range of particle sizes rather than a single species \citep{Krapp+etal2019,Paardekooper+etal2020,Paardekooper+etal2021}. 
In contrast, the growth rate and non-linear evolution for $\epsilon \gtrsim 1$ appear relatively unaffected by the size distribution \citep{ZhuYang2021,YangZhu2021,Schaffer+etal2021}. 
This implies that significant dust sedimentation is required before the streaming instability will be able to operate to form dense pebble filaments.

The strength of the turbulent diffusion in the protoplanetary disc is an important quantity that determines the efficiency of the streaming instability, since turbulence sets the scale height of the dust mid-plane layer \citep{Johansen+etal2014}. 
Background turbulence is also important because it can diminish the growth rate of the streaming instability or even entirely eliminate the growth of the instability  \citep{Umurhan+etal2020,ChenLin2020,Estrada+etal2022}. 
The dust scale height $H_{\rm p}$ has been observed to be 10\% of the gas scale height $H_{\rm g}$ in the outer regions of the well-resolved protoplanetary disc HL Tau \citep{Pinte+etal2016}. 
These observations imply a small dimensionless turbulent diffusion coefficient in the range of $\delta \sim 10^{-4}$ to $10^{-3}$. 
Recently, edge-on observations of another protoplanetary disc, Oph~163131, allowed estimates of a turbulent diffusion coefficient as low as $\delta \sim 10^{-5}$ \citep{Villenave+etal2022}. 
The turbulent diffusion coefficient can also be calculated from the radial width of pebble rings in protoplanetary discs \citep{Andrews2020}.
Here $\delta$ is inferred to be somewhat higher, $\delta \sim 10^{-3}$ \citep{Dullemond+etal2018}, but this is still within the range of weak turbulence. 
The gas turbulence exterior of planetary gaps can nevertheless be enhanced as a consequence of hydrodynamical instabilities that drive extra turbulence at a planetary gap edge \citep{Lyra+etal2009} and hence may not be representative of the bulk disc properties.

Full hydrodynamical modelling is needed to assess the co-evolution of the streaming instability together with other sources of turbulence (e.g., \citealt{Lesur2023}). 
The streaming instability has been studied together with both weak and strong turbulence driven by the magnetorotational instability, which can operate where the magnetic field pressure is not too high compared to the thermal pressure (i.e., the magnetic field, likely anchored in the giant molecular cloud, must be weak enough as to not repress the growth of the instability) and is sufficiently ionized that the magnetic field couples to the gas \citep{Balbus2003}. 
In the case of weak turbulence ($\delta \sim 10^{-3}$), the streaming instability grows from initial particle concentrations in low-amplitude pressure bumps that form in the turbulence by an inverse cascade of kinetic and magnetic energy \citep{Johansen+etal2007,Johansen+etal2009a}, while the case of strong turbulence ($\delta \sim 10^{-2}$) showed no evidence for feedback-regulated particle concentration \citep{Johansen+etal2011}. 
The magnetorotational instability is nevertheless likely to have been suppressed by the high resistivity to current in most parts of the protoplanetary disc \citep{Turner+etal2014,DeschTurner2015}. 
If the magnetorotational instability operates only in the upper layers of the disc, where the resistivity is lower, then density waves penetrating through the mid-plane stir the pebble layer to large-scale heights \citep{FromangPapaloizou2006}. 
Weak pressure bumps forming in the gas were nevertheless again observed to act as seeds for feedback-regulated particle concentration in such setups \citep{Yang+etal2018,XuBai2022}. 
The vertical shear instability is a prime candidate for driving turbulence in protoplanetary discs that does not depend on the resistivity level \citep{Nelson+etal2013}. 
The vertical shear instability evolved without particle feedback on the gas stirs up the particle mid-plane layer to high values \citep{Flock+etal2020,Schaefer+etal2020,SchaeferJohansen2022}. 
Including the particle feedback on the gas, the pebbles nevertheless sediment to a thin mid-plane, where the friction exerted on the gas suppresses the vertical shear instability \citep{Lin2019}. 
All in all, the streaming instability appears remarkably resilient to other sources of turbulence, when evolved together in hydrodynamical simulations, unless the turbulence reaches very high strengths of $\delta \sim 10^{-2}$.

The growth rate of the streaming instability is normally analyzed in a 2-D (radial-vertical) configuration and for a fixed mean dust-to-gas ratio \citep{YoudinGoodman2005,Umurhan+etal2020}, which ignores sedimentation of pebbles towards the mid-plane. 
Including this sedimentation in computer simulations has nevertheless demonstrated that the ratio of the pebble surface density to the gas surface density plays an important role in determining the non-linear outcome of the instability. 
Dense filaments only appear above a threshold metallicity \citep{Johansen+etal2009b,BaiStone2010}. 
This threshold depends strongly on the Stokes number St \citep{Carrera+etal2015,Yang+etal2017,LiYoudin2021} and on the strength of the background turbulence \citep{SchaeferJohansen2022,Lim+etal2023}. 
For a realistic Stokes number of 0.01, translating in the outer Solar System to a pebble size of 0.1--1 mm, the threshold is approximately 2\% metallicity, almost twice solar. 
For lower values of the metallicity, the streaming instability still develops turbulence that stirs up the mid-plane layer, but dense filaments do not form under such conditions. Increasing the local metallicity to trigger the conditions for planetesimal formation by the streaming instability may be possible by ice condensation near the water ice line \citep{RosJohansen2013,SchoonenbergOrmel2017,DrazkowskaAlibert2017,Ros+etal2019,Estrada+etal2023,RosJohansen2024} or by selective photoevaporation of the gas from the disc \citep{Carrera+etal2017,Ercolano+etal2017}.


\bigskip
\section{Initial Mass Function of Planetesimals from Streaming Instability Models} 

The formation of planetesimals with a characteristic size of a few hundred kilometers provided an early success of the streaming instability \citep{Johansen+etal2007}, given its correspondence to the typical size of asteroids in the main belt \citep{Morbidelli+etal2009}. 
Since then, much effort has gone into quantifying the initial mass function of planetesimals (which is related to, but not necessarily the same as, today's observed size distribution) formed by the streaming instability \citep{Johansen+etal2015,Simon+etal2016,Liu+etal2020,Schaefer+etal2017, Abod+etal2019, Li+etal2019}. 
Resolving the mass distribution below the characteristic mass requires computer simulations of very high resolution because of the multiscale nature of the pebble density field within the filaments formed by the streaming instability \citep{Johansen+etal2012}. 
The initial mass function appears to be a power law with the differential planetesimal number ${\rm d}N/{\rm d}M \propto M^{-1.6}$ \citep{Johansen+etal2015,Simon+etal2016}. 
The numerical value of the power law exponent is relatively independent of the physical parameters of the simulation \citep{Simon+etal2017}. 
Interpreting the mass function of planetesimals forming near the simulation resolution limit is nevertheless not straightforward, as a resulting lack of small planetesimals could either be due to insufficient resolution (and these small planetesimals would therefore appear once higher resolution simulations are performed) or due to an actual physical turn-over of the initial mass function for small masses dictated by the gravitational dynamics of pebbles moving within the turbulent flow \citep{Li+etal2019,KlahrSchreiber2020}.

Above the characteristic mass, the mass distribution in simulations appears to transition from a power-law to an exponential tapering (a transition that has also recently been observed in the size distribution of cold classical Trans-Neptunian Objects (TNOs; \citealt{Kavelaars+etal2021}). 
This transition makes physical sense, because the power-law exponent of $-1.6$ would otherwise integrate to infinite total mass. 
The largest planetesimal masses reach up to 100--1,000 times the characteristic mass \citep{Schaefer+etal2017,Liu+etal2019a}, i.e., corresponding to masses like those of Charon and Pluto, respectively, for typical simulation parameters. 
Simulations of the streaming instability on global disc scales will nevertheless be needed in the future to pin down the value of the most massive planetesimals, since the most massive planetesimals may form from rare high-density clumps that are statistically unlikely to form in small-domain boxes.
The most massive (and very rare) planetesimals residing at the end of the exponential tapering are also the most prone to continue to grow by pebble accretion  \citep{Liu+etal2019a,Lyra+etal2023}, with some contribution of planetesimal accretion only in the earliest growth decades \citep{LorekJohansen2022}, particularly in the inner regions of the protoplanetary disc. 
The mutual stirring between the growing protoplanets additionally helps to limit the number of protoplanets that grow to full planet mass \citep{Levison+etal2015}.

The characteristic mass of planetesimals formed by the streaming instability has been proposed by \cite{Liu+etal2020} to follow the simple scaling 
\begin{equation}
  M = 5 \times 10^{-5} M_{\rm E} \left( \frac{\Gamma}{\pi^{-1}} \right)^{a+1} \left( \frac{H_{\rm g}/r}{0.05} \right)^{3+b} \, .
  \label{eq:Mc}
\end{equation}
Here $M_{\rm E}$ is the mass of the Earth, $\Gamma=4 \pi G \rho_{\rm g}/\Omega^2$ (normalized here to $\Gamma=1/\pi$) is a parameter that measures the strength of the gas and particle self-gravity ($G$ is the gravitational constant, $\rho_{\rm g}$ is the mid-plane gas density, and $\Omega$ is the Keplerian orbital frequency), and $H_{\rm g}/r$ (normalized here to $H_{\rm g}/r=0.05$) is the ratio of the gas disc's scale height to the distance from the Sun. 
We ignored here the weak dependence on the global pressure gradient and on the local metallicity and take $a=0.5$ and $b=0$, following \cite{Liu+etal2020}. We refer to \cite{Abod+etal2019} for further discussion of these parameters. This expression is a simple consequence of a non-dimensionalization of the hydrodynamical equations that govern local computer simulations performed in the shearing box approximation. 
Other authors have proposed a characteristic mass that is partially derived from stability analysis of a self-gravitating protoplanetary disc \citep{Li+etal2019}; this yields a similar scaling with $H_{\rm g}/r$ as in equation (\ref{eq:Mc}) but with a much stronger dependence on $\Gamma$. 
Equation (\ref{eq:Mc}) can be transformed to give a characteristic diameter of
\begin{equation}
  d = 1000\,{\rm km} \left( \frac{\rho_\bullet}{500\,{\rm kg\,m^{-3}}} \right)^{1/3} \left( \frac{\Gamma}{0.3} \right)^{1/2} \left( \frac{H_{\rm g}/r}{0.05} \right) \, .
\end{equation}
A ``pristine'' protoplanetary disc with $\Sigma_{\rm g} \sim 10^3\,{\rm kg\,m^{-2}}$ in the trans-Neptunian region (so $\sim$$10^4\,{\rm kg\,m^{-2}}$ at 1 au) has $\Gamma \sim 1$. This value for the pristine surface density corresponds approximately to the transition to gravitational instability by the Toomre criterion; hence planetesimal formation at larger values of $\Sigma_{\rm g}$ would require consideration of strong background turbulence caused by the gravitational instability \citep{Gibbons+etal2015}.

\begin{figure}
  \begin{center}
    \includegraphics[width=1.0\linewidth]{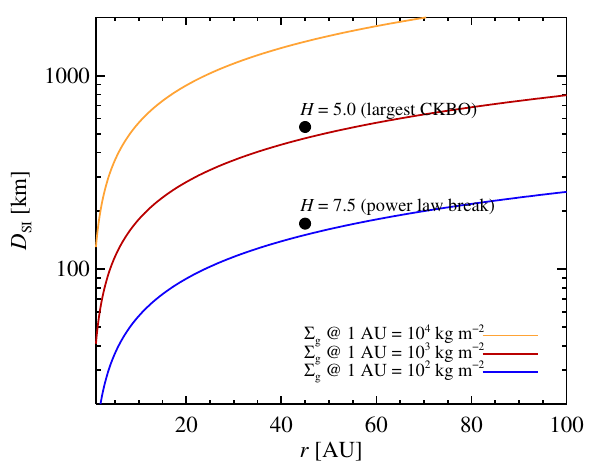}
  \end{center}
  \caption{The characteristic planetesimal diameter formed by the streaming instability, plotted as a function of the  distance from the star for three values of the gas column density at 1 au \citep{Liu+etal2020}. We also show the diameter conversion from the absolute magnitude $H$ of the largest objects among the cold classical TNOs (very low-inclination) population ($H=5$) and the fitted power law break ($H=7.5$; \citealt{Kavelaars+etal2021}). 
  These correspond well to the planetesimals formed by the streaming instability at the late stages of protoplanetary disc evolution, where the characteristic column density at 1 au has fallen from $10^4\,{\rm kg\,m^{-2}}$ to between $10^2\,{\rm kg\,m^{-2}}$ and $10^3\,{\rm kg\,m^{-2}}$.}
  \label{f:Rplanetesimal_r}
\end{figure}

In Figure \ref{f:Rplanetesimal_r} we show the characteristic planetesimal diameter from the streaming instability scaling as a function of distance from the star for three different evolution stages of the protoplanetary disc, assuming a low interior density of $500\,{\rm kg\,m^{-3}}$. We also indicate the largest object in the cold classical population as well as the characteristic size where the power-law size distribution for cold classical KBOs appears to transition to an exponential tapering \citep{Kavelaars+etal2021}; to transform from the observationally measured absolute magnitude $H$ to an estimated size, we assumed an albedo of $p=0.06$.
The cold classical population fits well with a formation in a protoplanetary disc that is depleted by a factor 10--100 relative to a pristine protoplanetary disc (which we consider here for simplicity to have approximately $\Sigma_{\rm g} \sim 10^4\,{\rm kg\,m^{-2}}$ at 1 au, with $1/r$ scaling with distance). 
Formation at such a late evolutionary stage of the protoplanetary disc is nevertheless still not consistent with the total mass of the cold classicals of only a few times $10^{-3}$ Earth masses \citep{Napier+etal2024}; even a narrow planetesimal formation region of 1 au width yields 10 times more mass than that for a depletion factor of 10. 
However, the scaling law for the planetesimal mass has not been tested for really low values of $\Gamma$ (e.g. 0.01, corresponding to 100 times depletion from a pristine protoplanetary disc) and it may well be that for very low $\Gamma$, the efficiency of turning pebbles into planetesimals is also low. Also, a steepening pressure gradient towards the outer regions of the solar protoplanetary disc would naturally lead to less efficient planetesimal formation \citep{BaiStone2010c,Abod+etal2019,Baronett+etal2024}.
Alternatively, the outward migration of Neptune by planetesimal scattering could have depleted the cold classical population by a factor 10 or more \citep{Gomes+etal2018}.

We therefore envision a formation of the classical Kuiper belt late in the evolution of the protoplanetary disc. 
Perhaps the outer disc lost gas mass by far-ultraviolet photoevaporation at this stage and this put the metallicity above the threshold \citep{Carrera+etal2017}. 
This late formation could also explain how the densities remained low \citep{Brown2013}, by avoiding heating by $^{26}$Al that melted the ice and rock of asteroids.  
\citep{Canas2024} explain the trend of the largest objects having higher densities ($\approx 2000$~kg/m$^3$) with a model in which the first planetesimals are primarily icy, but accrete silicate pebbles after $^{26}$Al has decayed. 

While the observed properties of the cold classical TNOs discussed above seem consistent with streaming instability formation models, the largest contributors to the Centaur population are the dynamically ``hot" TNOs, particularly the scattering TNOs (\citealt{Levison1997, Duncan1997, DiSisto2007, Volk2008, Nesvorny2017, Nesvorny2019, Disisto2020}; see also Chapter 3 by Di Sisto et al.).
The dynamically hot TNOs are generally assumed to have formed at smaller heliocentric distances ($r\lesssim30$~au) and transported outwards to their present-day orbits during the epoch of giant planet migration (see, e.g., \citealt{Gladman2021} for a recent review). The streaming instability scaling model presented in Figure \ref{f:Rplanetesimal_r} predicts that planetesimal birth diameters are significantly smaller in the 10--30 au region compared to the classical Kuiper Belt Objects. The large sizes present among the ``hot'' TNOs are consistent with formation closer to the Sun at earlier epochs of the protoplanetary disc where the gas and pebble column densities were larger.


\bigskip
\section{Observational Constraints for Likely Formation Regions of Modern Centaurs} 
\label{sec:formation_regions}

Clues to the histories of Centaurs may be found from
both telescopic surveys and from impact crater populations on outer Solar System bodies. 
Other useful observations for testing Solar System formation models include individual Centaur sizes, shapes and densities, colors, volatile and dust output, and levels of activity \citep{Fraser2022, Knight2023}. 
Here we highlight some key datasets and discuss some implications.

\subsection{Size Distributions of TNOs and Centaurs as Related to Formation Scenarios}

Constraints on the size distributions of the Centaurs and their source TNO populations come from a variety of observations. 
Optical telescopic surveys provide constraints for objects above $\sim 100$~km in diameter, with sizes inferred from the measurement of an object's absolute magnitude ($H$).
As discussed above, the cold classical TNO population is consistent with a power law size distribution that transitions to an exponential cutoff with no objects larger than $\sim400$~km \citep{Kavelaars+etal2021}.
The hot population that feeds the Centaurs has recently been found to share a very similar size distribution for the $\sim100$--400~km diameter size range \citep{Petit2023}, though it does not share the $\sim400$~km upper size limit with the cold population; all of the largest known TNOs, including the dwarf planets, belong to the hot population (e.g., \citealt{Schwamb2014}). 
Constraints on the smaller-end size distribution from TNO surveys are currently much weaker due to how faint smaller TNOs are at their large heliocentric distances.

Because the Centaur population is closer to the Sun than the TNOs, their size distribution can be measured to larger (i.e., fainter) $H$ magnitudes.
Their closer-in orbits also make them more amenable to more direct size measurements through a variety of techniques including thermal infrared measurements and stellar occultations. 
As of spring 2024, the sizes of only about five of the largest Centaurs have been measured by occultations with diameters ranging from $\sim$ 60 to 250 km (see Chapter 9 by Sickafoose et al.), providing some limited size overlap with the well-measured TNO size distribution.
Thermal infrared measurements have also been used to infer the diameters of dozens of Centaurs ranging down to $D\approx 20$~km \citep{Bauer2013}. 
For most Centaurs, however, only photometry at visual wavelengths is available. 
Assuming a geometric albedo of 0.08 \citep{Bauer2013}, the Centaur size distribution can be inferred for bodies as small as $d \approx$ 3--5 km, which is still larger than the well-measured portion of the Jupiter-family comet size distribution. 
See Chapter 4 by Fern\'andez et al. for a full discussion of constraints from the intermediate-sized and small Centaurs. 

Another source of information about Centaurs and their source populations for objects smaller than $\sim$100 km in diameter comes from the cratering records produced by their impacts on Pluto and Charon, the cold classical TNO Arrokoth, and the satellites of the giant planets. The giant planet satellites in theory should record the Centaur population and the (primarily) scattering TNOs from which they derive more directly than the Pluto system and Arrokoth, which interact with the full range of TNO populations. 
However, the giant planet satellite surfaces may also record planetocentric debris in some cases, and each satellite has its own unique geologic history which can complicate interpretations of the cratering record.
Thus, a comparison to the Pluto system and Arrokoth is still useful. 
Table~\ref{tab:craters} gives rough size ranges for the smallest and largest craters sampled in each giant planet system and for the worlds visited by New Horizons -- Pluto/Charon and Arrokoth. 

The smallest crater in each case is limited by image resolution and/or lack of image coverage and is not indicative of the actual smallest crater/impactor on the surface of each of these bodies. 
We use here the smallest crater sizes that are believed to be from primary impacts (i.e., not likely to be a secondary crater generated from ejecta of a primary impact), and that are well sampled (i.e., we did not consider them if there was just one very high-resolution image with one or a few craters in it). 
The representative largest craters noted here are features that are clearly visible today, and useful for understanding the shape of the size-frequency distribution up to a given diameter. 
Again, these are not necessarily the largest craters that have ever formed on these bodies, as those ancient impacts have been erased in many cases. 
Neptune's large moon Triton has a young and enigmatic surface \citep{Schenk2007, Mah2019}, so we do not include the Neptune system in this analysis.
The scaling to impactor sizes from crater sizes depends on the impactor and target properties, such as impact speed and body gravity, and is different for each body (e.g., \citealt{Holsapple1993, Housen2011}). 
Here we used the same methods as described in the supplement of \cite{Singer2019} for two different endmember materials, a ``solid'' ice surface and icy regolith, and reran the derivation for each satellite's average impact velocity \citep{Zahnle2003, Nesvorny2023} and surface gravity. 
We give the approximate size ranges of impactors recorded in each satellite system in Table~\ref{tab:craters}. 
The very smallest impactors are tens of meters in diameter, forming the smaller craters on the smaller satellites.  More typically, however, the impactors are hundreds of meters in diameter, up to tens of km for the largest impactors.

\begin{table}[!h]
\centering
  \begin{threeparttable}
\caption{\bf{Size ranges for craters and impactors for outer planet systems}}  \label{tab:craters}
\begin{tabular}{|m{5em}|m{8em}|m{8em}|m{7.5em}|m{7.5em}|}
 \hline
 \textbf{Location} & \textbf{Smallest crater size well sampled} & \textbf{Representative largest crater sizes} & \textbf{Impactor size range} & \textbf{Image sources} \\
 \hline
 Jupiter  &$\sim$1 km\tnote{1} & $\sim$150 km (several on Ganymede and Callisto)\tnote{2} \, and $\sim$570 km (Gilgamesh on Ganymede)\tnote{3} & $\sim$30 m to 10s of km ($\sim$50 km for Gilgamesh) & Galileo and Voyager 1 and 2 missions\\
\hline
Saturn (Mid-sized Satellites)&$\sim$1 km\tnote{4}&$\sim$300-600 km ($\sim$12 across the satellites with the largest being Turgis basin on Iapetus)\tnote{5}&
$\sim$10-70 m for the smaller craters and 10s of km for the larger basins (up to $\sim$70 km for Turgis)\tnote{6}&Cassini and Voyager 1 and 2 missions\\
\hline
Uranus&$\sim$1.5 km\tnote{7}&$\sim$100-340 km\tnote{7} 
 ($\sim$20 across the satellites, with the largest Gertrude basin on Titania)&10s of m to 10s of km (up to $\sim$40 km for Gertrude on Titania)&Voyager 2 mission\\
\hline
Pluto and Charon&$\sim$1.5 km\tnote{8}&$\sim$250 km (Burney basin)\tnote{8} and $\sim$1000 km (Sputnik basin)\tnote{9}&$\sim$150 m to 40 km\tnote{8} (for Burney Basin)&New Horizons mission\\
\hline
Arrokoth&$\sim$0.35 km\tnote{10}&$\sim$7 km (Sky crater)\tnote{10}&10s of m to $\sim$0.7 km\tnote{11}&New Horizons mission\\
\hline
\end{tabular}
    \begin{tablenotes}\footnotesize
      \item[1] The smallest and least eroded are on Europa: e.g., Bierhaus et al., 2009.
      \item[2] e.g., Schenk 2002.
      \item[3] Estimated equivalent diameter, Schenk et al. 2004.
      \item[4] This value is for some areas across most of the satellites e.g., Kirchoff and Schenk 2009, Kirchoff and Schenk 2010, Bierhaus et al. 2012.
      \item[5] Kirchoff et al. 2008, Kirchoff et al. 2018.
      \item[6] Range for this size of craters on various satellites and a range of target materials (from solid ice to icy regolith).
      \item[7] Kirchoff et al. 2022.
      \item[8] e.g., Singer et al. 2019.
      \item[9] e.g., McKinnon et al. 2017.
      \item[10] e.g., Spencer et al. 2020.
      \item[11] McKinnon et al. 2022.
    \end{tablenotes}
  \end{threeparttable}
\end{table}

\begin{figure}
    \centering
    \includegraphics[width=1\linewidth]{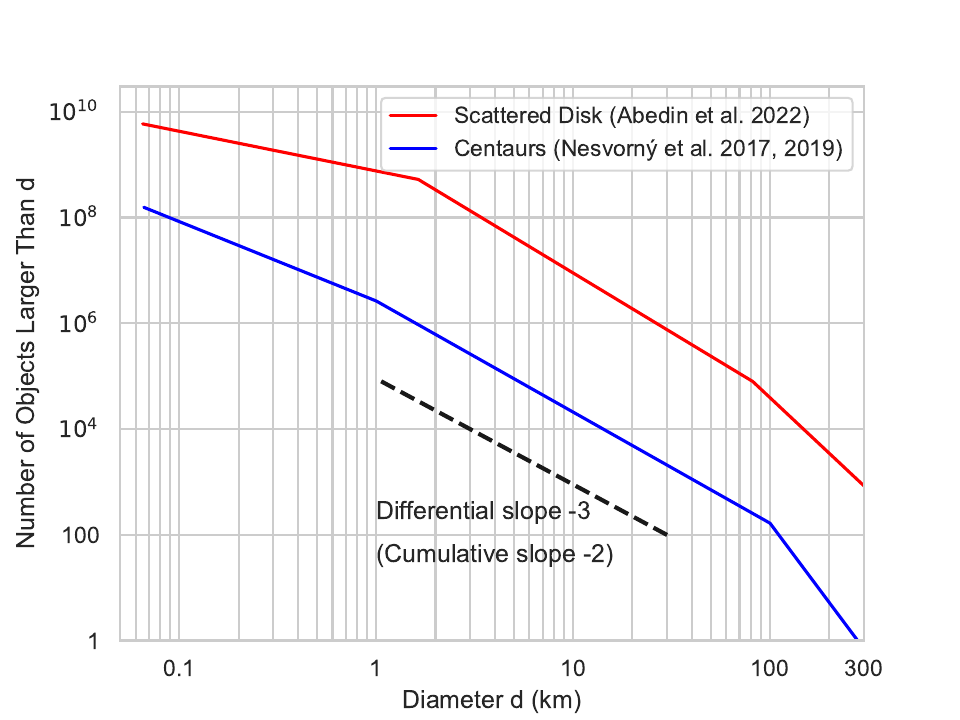}
    \caption
    {
    Schematic size distributions for scattered disk objects \citep{Abedin2022} and Centaurs \citep{Nesvorny2017,Nesvorny2019}. The size distributions are based on the cratering records of the giant planet satellites for diameters $d$ from $\sim 10$~m up to $\sim 70$~km, from
    telescopic observations of Centaurs for $d \gtrsim$ 3--5 km, and from telescopic surveys of TNOs for $d \gtrsim$ 20--50 km (see Table~\ref{tab:craters}). \cite{Abedin2022} assume there are $78,000 \pm 17,000$ ``scattering'' objects with absolute magnitude $H_r < 8.5$ and slopes $\alpha$ of the magnitude distribution of 0.7, 0.45, and 0.15 between $H_r$ values of 5.5--8.5, 8.5--17, and 17--24, respectively. 
    These values correspond to cumulative power-law size distributions with exponents $\gamma = 5\alpha$ equal to 3.5, 2.25, and 0.75 for diameter ranges of 328--82~km, 82--1.6~km, and 1.6--0.065~km, respectively, assuming a geometric albedo of 0.08. 
    \cite{Nesvorny2017} assume a cumulative size distribution with $\gamma$ equal to 5, 2.1, and 1.5 for diameter ranges of 300-100 km, 100--1~km, and less than 1 km, respectively, and an estimated population of $21,000 \pm 8,000$ Centaurs with diameters larger than 10 km, assuming a geometric albedo of 0.06. 
    The population of Centaurs at a given size is $\approx$~100--1000 times smaller than the number of scattered disk objects, due to the Centaurs' short dynamical lifetimes \citep{Irwin1995, Jewitt1996}.}
    \label{fig:centaur-sd-sfds}
\end{figure}

At the smallest sizes, the cratering record yields information on objects at least an order of magnitude smaller (and often two orders of magnitude smaller) than telescopic surveys of the TNO or Centaur populations. 
There is some overlap between the largest Centaur impactors recorded by craters and the smallest Centaurs and TNOs observed telescopically in the tens of km range. 
The largest craters are few in number, and the smallest  Centaurs are also difficult to observe, so there can be large uncertainties in both types of observations where the overlap occurs.
For ease of general comparison, here we summarize the crater size-distribution shapes in terms of one basic parameter: the average differential power-law slope over a given crater size range. 
A differential slope of -3 (-2 cumulative) is often seen and thus used as a reference as the ``typical'' or ``standard'' slope (see Figure~\ref{fig:centaur-sd-sfds}) that all others are either steeper than (e.g., -4) or shallower than (e.g., -2).
Additionally, most scaling theories would introduce a slight change in the size distribution slopes in the conversion from crater to impactors (e.g., as described in more detail in  \cite{Singer2021} for Pluto impactors). 
The impactor slope is slightly shallower than that of the craters. For comparison with the IMF obtained from streaming instabilities simulations discussed in Section 2.4, we note that a cumulative power law in diameter $N_> \propto d^q$ corresponds to a differential power law in mass $dN/dM \propto M^{q/3-1}$, so that $q=-2$ yields $q'=-5/3$ in terms of mass.

The cratering record at Pluto and Charon primarily records impacts from the hot classical TNO population and the rest of the Plutino population rather than from the scattering population \citep{Greenstreet2015, Greenstreet2016, Greenstreet2023}.
While this means it is not directly probing the dominant Centaur reservoir, the hot classical and resonant TNO populations were likely sourced from a similar region of the planetesimal disk as the scattering TNOs.
The largest craters on Pluto/Charon (\textgreater10 km craters, or \textgreater1 km impactors) display a -3 average slope, similar to the slope inferred from optical surveys for TNOs smaller than about 100~km \citep{Bernstein2004, Fraser2014, Greenstreet2015, Greenstreet2016}. 
There is a distinct break to a shallower slope (approximately -1.7) for smaller craters (\textless10 km craters, or \textless1 km impactors) on a wide range of geologic terrains \citep{Singer2019, Singer2021}. 
The impacting population at Arrokoth is dominated by the moderately dynamically excited `stirred' classical TNOs (see full discussion in \cite{Greenstreet2019}.
All of the measured craters at Arrokoth are made by impactors less than $\approx$~1~km in diameter and display a shallow slope consistent with the Pluto/Charon record \citep{Spencer2020, McKinnon2022}.

The main impacting small bodies on the giant planet satellites are Jupiter-family comets, Centaurs, and, in the case of Neptune's satellites, the scattered disk objects themselves \citep{Zahnle2003, Nesvorny2023}.
In the Jupiter system, the surfaces where smaller primary craters can be observed, are generally the young surfaces on Europa and Ganymede.
These terrains show a similar trend to that seen on Pluto and Charon, where the more typical average -3 slope is seen for larger craters, and there is a transition to a similar shallower slope (-2 or shallower) below craters that correspond to an impactor diameter of $\approx$~1~km \citep{Schenk2004, Bierhaus2009, Singer2019}.
The crater distributions at Saturn generally show ``typical'' or somewhat steeper slopes at large sizes ($>$~a few km) and become shallower at smaller sizes \citep{Kirchoff2009, Kirchoff2010, Bierhaus2012, Robbins2024}. 
These shallow craters occur at impactor sizes of tens to a few hundred meters, which is about an order of magnitude smaller than the $\sim1$~km transition size in the Pluto/Charon system.
However, the Saturn system is complex, with different geologic histories for the inner vs. outer satellites, the potential for planetocentric crater populations, and ongoing discussions of the origin and overall age of the satellites  \citep{Dones2009, Kirchoff2018}. 
Thus, it is harder to know what portion of the craters in the Saturn system is representative of the heliocentric Centaur impactor population.
Miranda is the uranian satellite where craters smaller than $\approx$~10~km can best be mapped \citep{Strom1987, Plescia1988, Kirchoff2022}. 
Different terrains across Miranda show different size-distribution slopes. The older cratered terrains show a break to a shallow slope similar to that on Pluto/Charon, but at a slightly smaller corresponding impactor size of $\approx$~300--500~m. 
However, the craters superposing younger tectonized terrain remain closer to the typical -3 slope even at small crater sizes. 
The reason for these differences with terrain age are not known, but given the similarity of the younger terrains to Pluto/Charon, they may better represent the heliocentric impactor population  \citep{Kirchoff2022}.


Many of these surfaces in the outer Solar System are young and show no obvious signs of geologic processes preferentially erasing small craters (e.g., \citealt{Singer2019, Bottke2024}). This means the craters record the more recent impact flux with relatively fewer small impactors.
The shallower slope at small impactor sizes seen on many surfaces across the outer Solar System could indicate that smaller objects were not made in large abundances during either their formation or subsequent collisional evolution. 
The alternative would be that many small Centaurs and scattered disk objects were made throughout Solar System history, but somehow the smaller ones were preferentially removed before impacting these surfaces. 
The latter scenario seems difficult, given there are currently no known processes that would preferentially remove smaller objects, especially across many TNO sub-populations and from $\approx$~5 to 50 au. 
Disruption of small comets with perihelion distances $q_p \lesssim$~2~au does proceed rapidly \citep{Jewitt2021}, but this appears to result from spin-up due to outgassing torques. 
All known active Centaurs have $q_p < 12$~au \citep{Jewitt2009, Lilly2024}, so it seems unlikely that small objects can be removed all the way out to 50~au. 

If the deficit of objects below about 1 km in diameter seen in the crater records indicates that not as many of those objects formed originally, this could be consistent with direct formation of large Kuiper Belt objects via streaming instabilities and gravitational collapse \citep{Singer2019} rather than by hierarchical growth from small bodies \citep{Schlichting+etal2013}. For the streaming instability IMF to be consistent with the apparent lack of small planetesimals, this seems to require also a lower-mass cut-off to the bodies that can form by the streaming instability, maybe caused by diffusive outwashing of the smallest pebble clumps \citep{KlahrSchreiber2020}.

\subsection{Spectra and Colors of Centaurs and other Relevant Bodies}

Observational clues to chemical compositions and physical conditions may come from icy and refractory abundances measured on the surfaces of inactive Centaurs and in the comae of active Centaurs. 
For example, some interesting differences were recently found using millimeter- and infrared wavelength spectroscopy of CO and CO$_2$ produced in the coma of a highly processed Centaur (39P/Oterma) and one that has yet to enter the inner ($<$ 4 au) Solar System (29P/Schwassmann-Wachmann), but volatile species have been detected in only a few Centaurs, so it is difficult to draw broader conclusions about Centaurs' chemical compositions \citep{Womack2017, Wierzchos2017, Bockelee-Morvan2022, HarringtonPinto2023}.  
The relative gas abundance ratio of the highly volatile species CO and CO$_2$ is high in 29P but much lower in 39P  and this may be partly attributed to the objects' very different orbital histories and thus the different amounts of solar radiation processing they experienced \citep{HarringtonPinto2022, HarringtonPinto2023}. Measurements of the relative abundances of CO, CO$_2$, and other key species are needed in many more Centaurs in order to put tight constraints on formation regions as well as physical evolution models and is addressed more in Chapter 6 by Mandt et al. 

Unfortunately, most Centaurs are too small and too far from the Sun to readily obtain useful spectra for any but the brightest objects. When objects are too faint for spectroscopic techniques, an alternative way to gain insight is by analyzing their ``colors,'' which are calculated from the measured magnitudes of objects as seen through narrowband or broadband filters. Because this technique collects much more light than most spectra, one can obtain data for much fainter objects.  
Color diagrams can be used to analyze large-scale surface and gaseous properties of larger ensembles. This allows for an improved description of the Centaur population and can serve as the basis for statistically significant tests for correlations with other observed characteristics and can constrain models of Centaur and TNO formation, which we briefly discuss.

The first two Centaurs to be identified, 2060 Chiron and 5145 Pholus, span almost the full range of optical colors seen at visual wavelengths for outer Solar System bodies. 
Chiron's nucleus surface is a nearly gray, or neutral, reflector (e.g., B-R $\sim$ 1), while Pholus' is very red (B-R $\sim$ 2), once again proving that it is impossible to assess a new population with only two members. Initial Centaur surveys indicated that their colors had a bimodal distribution, but it was unclear whether the two color populations were the result of (1) evolutionary processes such as radiation-reddening, collisions, and sublimation or (2) a primordial, temperature-induced, composition gradient \citep{Peixinho2003, Tegler2008, Peixinho2012}. 

Intercomparing the colors of Centaurs with those of hot TNO/KBO populations can provide strong diagnostics for models \citep{Brown2011}. 
Interestingly, one of the first color magnitude surveys of Kuiper Belt Objects also found evidence for two distinct populations.
\citep{Doressoundiram2002, Doressoundiram2003, Trujillo2002} found that most cold classicals were very red, while ``hot'' classicals had a wide range of colors. 
To explain both the dynamically un-excited orbits of the cold classicals as well as their dominantly red surface colors, they proposed, following \cite{Levison2001}, that the cold classicals formed {\em in situ}, while the other Kuiper Belt populations had been implanted from a region of the protoplanetary disk interior to the present-day belt during planetary migration (e.g., \citealt{Malhotra93}). 
Outward migration of the hot TNO populations was elaborated on in the ``Nice model'' \citep{Tsiganis05, Morbidelli2005, Gomes2005} and subsequent models in which the giant planets underwent an orbital instability (see reviews by \citealt{Dones2015} and \citealt{Nesvorny2018}).
The different proposed migration histories of the giant planets have important implications for the original formation locations of the hot TNOs that feed the present-day Centaur population which may be reflected in their surface colors.

\begin{figure}
    \centering
    \includegraphics[width=1\linewidth]{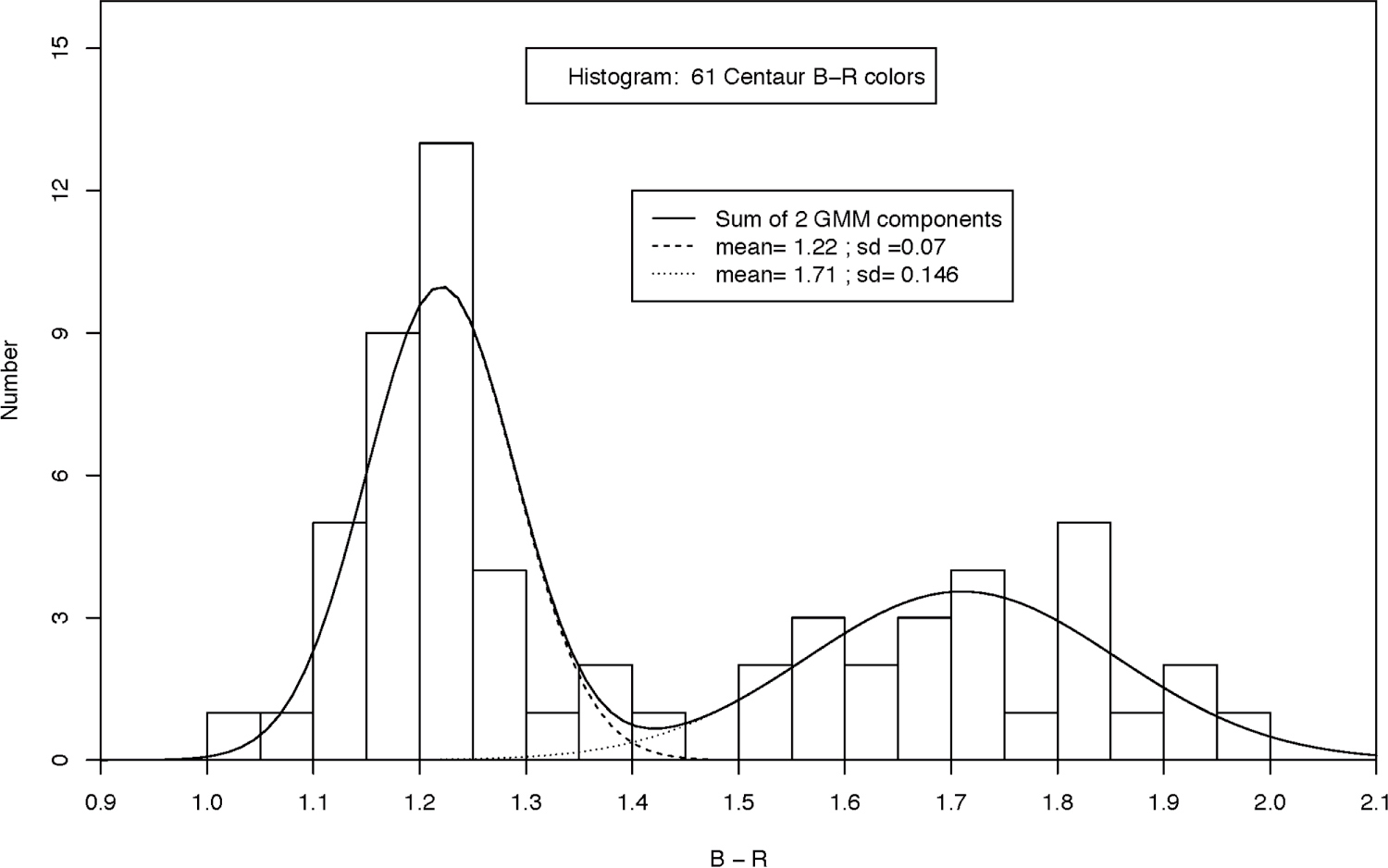}
    \caption{The distribution of $B - R$ colors measured for 61 Centaurs, from \cite{Tegler2016}. The color distribution is not bimodal at a statistically significant level, although a sample size of two or three times larger may be necessary to further test for bimodality. Including this sample, a bimodal color distribution is seen for intrinsically faint (small) and intrinsically bright (large) Centaurs and TNOs, consistent with Peixinho et al. (2012).}
    \label{fig:tegler2016}
\end{figure}

One important test will be to see whether the colors of a large sample of Centaurs are consistent with hot TNO populations \citep{Brown2011}. 
\cite{Peixinho2003} initially argued that Centaurs have a bimodal color distribution and TNOs do not. 
\cite{Peixinho2012} found that both Centaurs and {\em small} TNOs ($H_r \geq 6.8$, corresponding to diameters $\lesssim 165$~km for an assumed geometric albedo of 0.09) had bimodal color distributions, i.e., the bimodality depended on size and not dynamical classification. \cite{Wong2017} confirmed this result for small TNOs. 
However, Figure~\ref{fig:tegler2016} shows the distribution of $B - R$ colors measured by \cite{Tegler2016} for a larger survey of 61 Centaurs (twice as many as in the Peixinho analysis). 
The red peak is broader than it was for the smaller samples of objects in earlier studies with smaller sample sizes, and the color distribution is not bimodal at a statistically significant level. 
Nonetheless, the greater prevalence of very red Centaurs, as compared with main-belt asteroids \citep{Hasegawa2021}, Jupiter Trojans, and Hildas \citep{Wong2015, Wong2017-Hildas-Trojans} still needs an explanation. Interestingly, no Jupiter-family comets are found with ultra-red surfaces that are commonly seen in Centaurs; some have attributed this difference to the cometary activity of JFCs \citep{Jewitt2002, Grundy2009}. 
\cite{Melita2012} added that many gray Centaurs likely had cometary activity and a more detailed comparison between active Centaurs and Jupiter Family comets will be useful for constraining models. 

Many recent works argue that planetesimals in the outer Solar System formed with a range of compositions, but with little variation with distance from the Sun. Others propose that the range of colors seen today could result from distance-dependent sublimation of ices due to solar heating \citep{Davidsson2021, Parhi2023}, followed by irradiation, which we describe below. 
For example, \cite{Brown2011} proposed that in the innermost primordial Kuiper Belt, water, carbon dioxide, and possibly hydrogen sulfide are the only thermally stable ices. 
Irradiation would chemically process this mixture into a low-albedo, gray substance. 
Beyond about 20~au, methanol is also stable, and irradiation would yield red planetesimals with somewhat higher albedos. 
Finally, beyond about 35~au, ammonia would also be stable, irradiation of which would (hopefully) produce the unique properties of cold classicals. Thus, color measurements, as well as spectroscopic studies, can provide important tests for these models. 

Compositional information of Centaur nuclei and comae are expected to be useful to tie back to the formation location in the disk, and thus the conditions for the streaming instability. Additional discussion about Centaur surface temperatures (e.g., \citealt{Luu1993, Campins1994}) and compositions obtained from optical and infrared spectra of Centaur surfaces is provided in more detail in Chapter 5 by Peixinho et al. With the operation of JWST and other next-generation optical telescopes, many Centaur observing programs are underway which will very likely provide unprecedented spectra for many more Centaurs, as discussed in Chapter 13 by 
Fern\'andez-Valenzuela et al.

\subsection{Additional Useful Observational Constraints}

Other vital information for constraining Solar System formation models may be obtained from observations of 
Centaurs' shapes, densities, and the prevalence of binaries and/or have moons. We briefly consider some of these characteristics here. 

We are not aware of specific predictions for Centaurs, although \cite{Fraser2017} suggested that all planetesimals born near the Kuiper belt formed as binaries and \cite{Nesvorny2019-SI} 
asserted that wide TNO binaries provided evidence for the streaming instability. Thus, the extent to which Centaur binaries may exist is also likely to be an important modeling constraint. Not much is known about how many binary Centaurs exist, although the planet-crossing objects 42355 Typhon/Echidna (semi-major axis $a = 37.5$~au, perihelion distance $q_p = 17.5$~au) and 65489 Ceto/Phorcys ($a = 99.2$~au, $q_p = 17.7$~au) are considered to be Centaur binaries for some extended definitions of the centaur population \citep{Araujo2018, Grundy2007}. \cite{Santos-Sanz2012} infer low densities for these systems -- $360^{+80}_{-70}$ and $640^{+160}_{-130}$~kg/m$^3$, respectively. 
The masses of these systems are known to better than 10\%, but the sizes of the objects are less certain. 
For comparison, \cite{Berthier2020} infer a density of $810 \pm 160$~kg/m$^3$ for the jovian Trojan binary 617 Patroclus/Menoetius, whose mass is comparable to the masses of the Typhon/Echidna and Ceto/Phorcys systems, while \cite{Carry2023} find a density of $830 \pm 50$~kg/m$^3$ for the less massive bilobate Trojan 17365 Thymbraeus; see \cite{Mottola2024} for a review.


Centaur shapes are still poorly known. Interestingly, many TNOs and JFCs with well-determined shapes (from spacecraft encounters and radar) have two lobes. It is unknown whether all of these bilobate shapes are primordial or due to physical evolution \citep{Hirabayashi2016, Steckloff2016, Schwartz2018, Vavilov2019, Safrit2021,LorekJohansen2024}, however, the two-lobed structure of the cold classical TNO Arrokoth is believed to be from its initial formation \citep{McKinnon+etal2020,Spencer2020,Lyra+etal2021}, not later breakup and re-accretion as many asteroids binaries are thought to be.  Given that Centaurs represent the transitional state between TNOs and JFCs, it would not be surprising to find some bilobate Centaurs. None have been detected thus far.



Observations like these and many others may also help us determine which size bodies are primordial, rather than collisional fragments  \citep{Morbidelli2015, Benavidez2022, Jutzi2017, Davidsson2016}.
Also, survival of hypervolatiles in catastrophic collisions in the early Solar System has also been addressed (cf. \citealt{Davidsson2023}) and has implications  for Centaurs.


\bigskip
\section{Future Priorities for Understanding Planetesimal Formation in the Solar System}

Planetesimal formation theory has benefited enormously from observations, both from the ground and with dedicated satellite missions, of asteroids, Centaurs, and Kuiper Belt Objects. In particular, the size distributions and binary, and shape properties of planetesimals in the Solar System provide excellent measures to test against computer simulations of planetesimal formation. Many aspects of planetesimal formation by the streaming instability are now relatively well-understood (such as the binary properties and the characteristic masses), while others will need significant additional work to fully map (such as the dependence on the background turbulence and the timing of planetesimal formation in the protoplanetary disc). Probing the initial mass function of planetesimals down to much smaller diameters (10 km or even 1 km) will require computer simulations that resolve all the relevant scales of particle concentration and the gravitational collapse. Adaptive mesh refinement methods, such as used in \cite{Schaefer+etal2020}, may be a key technology to exploit to reach effective higher resolutions without the necessity of excessive computational resources.

Similarly, measuring the Centaur and TNO source population size distributions down to smaller sizes with telescopic surveys may provide critical constraints on planetesimal formation.  Recent and future Earth- and space-based telescopes such as the James Webb Space Telescope, the Vera Rubin Observatory, and the Nancy Grace Roman Space Telescope will greatly assist in understanding the size distribution of these objects below 50 km, and the latter can even push well into the sub-km regime.  Greater sampling of the spectra and color of Centaurs and TNOs will also greatly enhance our statistical understanding of their compositions and also help us understand how much post-formation processing these objects have experienced.

\bibliographystyle{apalike}
\bibliography{bibliography}

\end{document}